# Social groups in pedestrian crowds: Review of their influence on the dynamics and their modelling

Alexandre Nicolas[a] and Fadratul Hafinaz Hassan[b]

[a] Institut Lumière Matière, CNRS & Université Claude Bernard Lyon , F-69622, Villeurbanne, France

[b] School of Computer Sciences, Universiti Sains Malaysia, Pulau Pinang 11800, Malaysia

Correspondence: Alexandre Nicolas, alexandre.nicolas@polytechnique.edu and Fadratul Hafinaz Hassan, fadratul@usm.my

**Abstract**

Pedestrians are often encountered walking in the company of some social relations, rather than alone. The social groups thus formed, in variable proportions depending on the context, are not randomly organised but exhibit distinct features, such as the well-known tendency of 3-member groups to be arranged in a V-shape. The existence of group structures is thus likely to impact the collective dynamics of the crowd, possibly in a critical way when emergency situations are considered. After turning a blind eye to these group aspects for years, endeavours to model groups in crowd simulation software have thrived in the past decades. This fairly short review opens on a description of their empirical characteristics and their impact on the global flow. Then, it aims to offer a pedagogical discussion of the main strategies to model such groups, within different types of models, in order to provide guidance for prospective modellers.

**Introduction**

Pedestrians are the atoms of moving crowds, that is – etymologically speaking – their indivisible constituents. But every physicist or chemist knows that the properties of matter are not determined solely by the atomic content: The numerous facets of carbon are cogent evidence that the arrangement of these atoms and the molecules that they may form are also of great significance. Pedestrians, too, are often found in 'molecular' form (McPhail & Wohlstein, 1982; Zhou et al., 2017), i.e., clustered into groups of related people walking together. Still, the dynamics of crowds were long studied and modelled as though they consisted of single individuals. The effects of the interactions between relatives, friends, or colleagues were thus disregarded to a large extent (Huang et al., 2018).

The past decade, however, witnessed a Boylean[1] evolution of the field, with a growing number of studies dedicated to the influence of social groups on pedestrian dynamics. At the 'microscopic' level, it was empirically found that, far from being amorphous aggregates, social groups of pedestrians tend to adopt particular shapes (Moussaid et al., 2010) and to move slower than single pedestrians. At the collective scale, the impact of these groups on major flow properties - in particular on the fundamental diagram that relates the pedestrian density and the flow rate (Hu et al., 2021) - and on evacuation times in the event of an emergency (Ma et al., 2017; Li et al., 2020) has also received considerable attention, but the observed effects are either less marked or more controversial. These questions are however of paramount importance to properly dimension facilities with different types of crowd compositions or to estimate evacuation times of crowded rooms, and have recently seeped into new scientific areas in the context of the Covid-19 pandemic, which prompted physical distancing guidelines (Pouw et al., 2020; Garcia et al., 2020). Naturally, these empirical endeavours are mirrored by similar efforts in numerical modelling in order to properly integrate social groups in the simulations; solutions must be put forward and adapted to the types of models on the market.

This short review does not purport to provide a comprehensive account of the already bulky body of literature on social groups. Our goal is rather to shed light on the prominent characteristics and effects of social groups that have been established empirically, for the non-expert and expert readers alike, and to delineate the main strategies that have been put forward to model such groups, with their successes and limitations. By focusing our explanations on an inexhaustive list of papers, we hope that the manuscript can be somehow more helpful than a shallow enumeration of past works, but no efforts will be made to delve into the technical details of each work, which are best explained in the associated manuscript.

After defining social groups and exposing their microscopic characteristics in **Chapter I**, we will briefly review the effects of such groups on the collective flow properties in **Chapter II**. Finally, **Chapter III** will be dedicated to the main strategies to implement social groups (in terms of interactions among group members, but also between groups) in discrete models as well as in continuous ones.

---

[1] Robert Boyle (1627-1691) postulated that matter is composed of clusters of particles whose arrangement is responsible for its chemical properties.

# I. Social groups of pedestrians: definition, prevalence, and empirical characteristics

## *A. Definition, prevalence, and size of pedestrian groups*

In an interdisciplinary field that gathers researchers with very different backgrounds, it comes as a pleasant surprise that there appears to be a broad consensus regarding what social groups in pedestrian crowds are and how to define them. In particular, to ward off the risk of an all-too-loose definition whereby the whole crowd would be regarded as a single group, emphasis is generally put on the need for social ties (of any type) between group members. For instance, (McPhail and Wohlstein, 1982), cited and followed by (Zanlungo et al., 2014), require that group members be '*engaged in a social relation to one or more pedestrians and move together towards a common goal*', while they are viewed by (Moussaid et al., 2010) as '*individuals who have social ties and intentionally walk together, such as friends or family members*', a wording largely shared by (Huang et al., 2018). More simply, Stangor defines them as '*three or more people who are perceived, by themselves or others, to be a group*' – note that the lower bound of three people (triad) has not been taken up in the field of pedestrian dynamics: Dyads (groups of two people) are generally accepted among social groups.

Once thus defined, the proportion of such groups (as opposed to individuals walking alone) strongly varies with the context of observation. In **Table 1**, we have summarised the distributions of group sizes (quantified by the number of *people* in each category) found in diverse crowds, following published reports. Clearly, depending on the context, the proportion of single pedestrians may range from as low as a few percent, for instance at popular Fairs in Germany (Oberhagemann, 2012), up to more than 80%, in an underground pedestrian facility in Japan (Zanlungo et al., 2014). The time of the day or period in the year may also entail changes, with presumably fewer groups on workdays and at rush hours. Accordingly, it makes little sense to reason in terms of a context-independent average proportion of social groups; one had better aspire to understand how these groups impact the collective dynamics as a function of their prevalence in each context.

Table 1 – Statistical distribution of groups as a function of the group size, i.e., the estimated number of members. The given percentages refer to the number of *people* in each category.

| Group size = | 1 | 2 | 3 | 4 | 5+ |
|---|---|---|---|---|---|
| Nottingham train station (Singh and Drury, 2009) | 44% | 36% | 10% | 5% | 4% |
| Nottingham University campus (Singh and Drury, 2009) | 53% | 28% | 15% | 2% | 2% |
| Fairs in Germany (Oberhagemann, 2012) | 3% | 51% | 19% | 17% | 9% |
| Sport events [unpublished work by McPhail, 2003, cited by (Ge et al., 2012)] | 11% | 37% | 20% | 32% | |
| Sport event (Ge et al., 2012) | ~50% | ~30% | ~20% | – | – |
| University campus (admission test), Italy (Federici et al., 2012) | 34% | 44% | 17% | ~5% | |
| Hospital (You et al., 2016)[2] | 20% | 41% | 17% | 23% | |
| City square (You et al., 2016) | 19% | 22% | 25% | 33% | |
| Canteen (You et al., 2016) | 31% | 25% | 16% | 28% | |
| Commercial street (Crociani et al., 2013) | 16% | 31% | 18% | ~36% | |
| Underground pedestrian facility in Osaka, Japan (Zanlungo et al., 2014) | 82% | 15% | 3% | <1% | |
| Protestant Church Day in Germany (Schultz et al., 2014) | 14% | 41% | 17% | 12% | 15% |
| Commercial street in Toulouse (low density) (Moussaid et al., 2010) | 46% | 37% | 10% | 4% | 3% |
| Commercial street in Toulouse (moderate density) (Moussaid et al., 2010) | 29% | 51% | 12% | 5% | 3% |

The definitions in the foregoing paragraphs highlight that sheer proximity does not turn any cluster of people into a social group; this comes as a distant echo to Freud's assertion that not all crowds are psychological crowds (Freud, 1955). And, indeed, this restrictive definition generally makes sense as far as one is concerned with the influence of such groups on the dynamics of the crowd. However, in some situations, especially in cases of emergency, groups may spontaneously unite neighbouring individuals without prior social ties. For instance, in an evacuation, people may assemble to look for an exit or take joint action. Such groups are referred to as *physical groups* in (Ta et al., 2017). The ties linking their members are presumably much weaker than those binding parents and children, or friends, with direct consequences on the behavioural choices in these emergency conditions. More generally, it is easy to understand that, similarly to bonds between atoms, the ties binding group members are of variable strengths, ranging from loose connections between colleagues to hand-holding and tight embraces.

---
[2] Regarding this work, there remains an uncertainty as to whether the proportions relate to the number of pedestrians or to the number of groups.

## *B. Characteristic group shapes*

Since the pioneering works of E. Hall (Hall et al., 1968) on proxemics, it has been known that the short-range interactions between people follow some rules, for instance related to the preservation of one's private space, which may vary between cultures. It is now clear that such rules do not exist only for static interactions, but also for moving groups of pedestrians: Typical patterns have been brought to light in the shapes of small groups (up to 5 people) and relative positions of their members, whereas larger groups are likely to split into more stable small subgroups (Costa, 2010). Indeed, when group members are too far away from each other to communicate, they may break off and only consider those in the immediate surrounding as the current social group; the rate at which members are lost may then be proportional to the group size (Moussaïd et al., 2010). Along the same lines, (Zanlungo and Kanda, 2013) found that the social interaction between members weakens for groups comprising more than two people.

Regarding small groups, striking results have consistently emerged about their shapes: Members of a dyad generally walk abreast (i.e., forming a line perpendicular to the direction of walking), with a spacing distance around 80 cm from centroid to centroid (Moussaid et al., 2010; Zanlungo et al., 2014). Triads also walk more or less abreast, but with a significant bend in the middle of the line: a **V**-shape, with the person in the middle standing 20 cm or more behind the side members, was evidenced by the empirical observations of (Costa, 2010) as well as those of (Moussaïd et al., 2010) in a commercial street in Toulouse, France, and later confirmed in several other works, in a student union building, at a Church Congress in Germany, in an underground pedestrian facility in Japan, etc. (Ge et al., 2012; Zanlungo et al., 2014; Schutz et al., 2014). In sparse settings, groups of size 4 and 5 also tend to walk more or less abreast, with the central members slightly backwards compared to the two side members (Zanlungo et al., 2014; Schutz et al., 2014). This corresponds to a **U**-shape. Note, however, that other shapes were also observed by (Federici et al., 2012) on the campus of University of Milan, Italy; close to one third of the four-member groups were reported to be in a rhombus-like shape. In **Figure 1**, we have gathered empirical measurements of the relative positions of pedestrians in groups of different sizes and adjusted their dimensions so that they have the same scale; this comparison bolsters the consistency of the reports about typical group shapes.

**Figure 1**: Empirical statistical distributions of the relative positions of pedestrians in groups of distinct sizes, **(a)** at a Church Congress in Germany, **(b-c)** on a commercial street in France, **(d)** in an underground pedestrian facility in Japan. The figures were adapted from the reference at the left of each row and adjusted to the same scale.

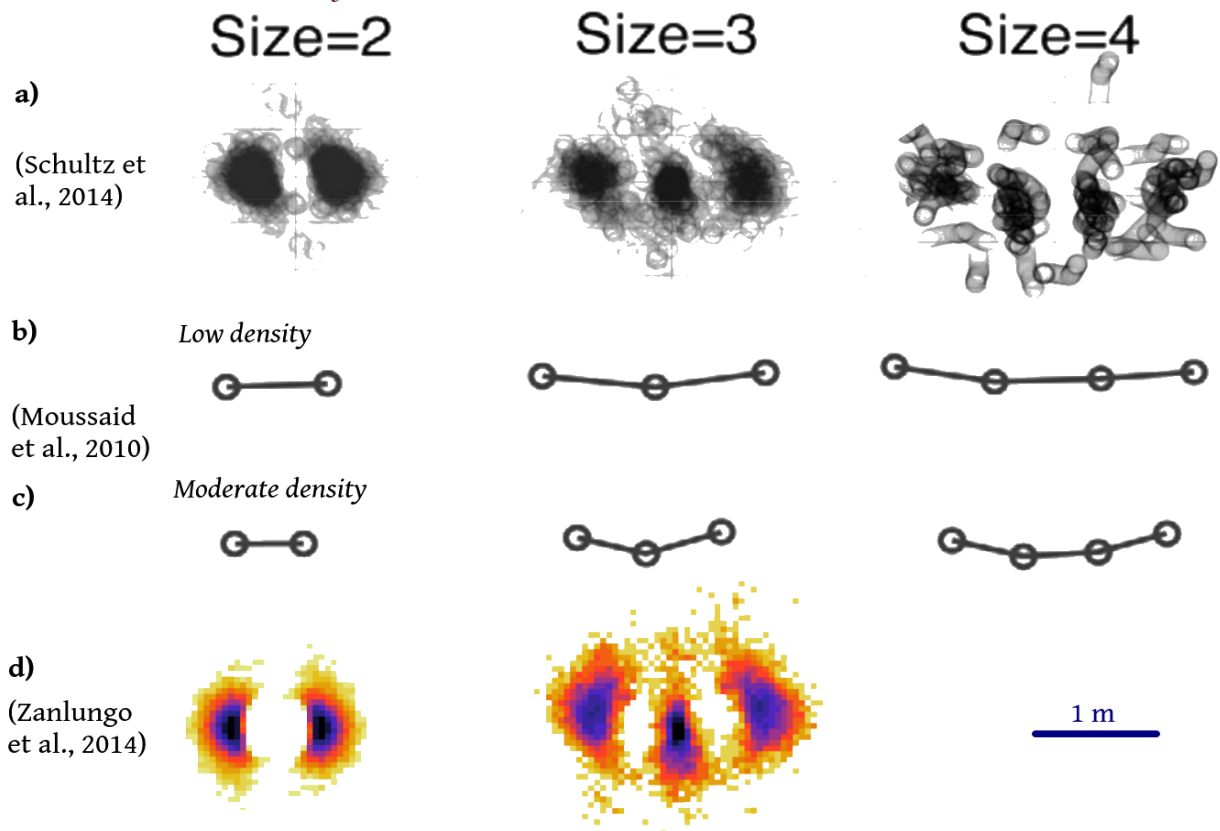

It is arresting that these group shapes stand in stark contrast with what one would expect from aerodynamics. Indeed, aerodynamics would favour $\Lambda$-shapes (i.e., inverse **V**-shapes), with a prominent leading edge to deflect air flows (think about the nose of an airplane). Similar $\Lambda$-shapes make the front leader more visible to other members, which facilitates following behaviour; this $\Lambda$-shape is thus observed in bird flocks. In contrast, the **V** and **U**-shapes observed in pedestrian groups are likely to originate from its making communication easier (Moussaïd et al., 2010) and requiring less twisting of the head (whose most comfortable orientation is in the walking direction) to gaze at the closest neighbour(s) (Zanlungo et al., 2014).

Nevertheless, these shapes are subject to variations. Notably, when the crowd's density increases, the configuration of the group becomes more constrained and it tends to shrink laterally, as shown in **Fig. 1(c)**. Pushed to an extreme, this leads to the formation of river-like groups, reminiscent of ant columns (Helbing et al., 2005), even though the analogy between pedestrians and ants is arguable. As a matter of fact, small groups naturally exhibit characteristic modes of fluctuations, revealed by principal component analysis (Ge et al., 2012); these include a lateral stretching mode and a streamwise elongational ('deformation') mode. This structural flexibility of the group enables it to adapt to external constraints. Finally, Zanlungo and colleagues have paid particular attention over the past years to the factors that may influence the configurations of small groups. Among other results, they observed that the relation binding group members affects the structure, with couples walking very close and abreast while dyads and triads of colleagues (work-oriented groups) keep a larger spacing distance (Zanlungo et al., 2017; Zanlungo et al., 2019). Regarding gender, (Costa, 2010) reported a stronger tendency for abreast walking in mixed dyads and then female dyads, compared to male dyads or triads, whereas same-gender triads were found to be more ordered than mixed ones by (Zanlungo et al., 2019). Besides, men within a triad tend to keep more distance between themselves than women (Zanlungo et al., 2019). All in all, while there are robust results concerning group shapes, various factors such as gender, purpose, and personal relation within the group influence the detailed group configuration.

## C. Group speeds

Beyond the shape of groups, robust findings have also appeared about their speed. In particular, field studies have consistently shown that groups move slower than single pedestrians and that their speed is reduced as the group size increases, up to 3 or 4 members, at least in not too dense settings. For instance, (Gorrini et al., 2016) measured that dyads walk 30% more slowly than single individuals in the pedestrian gallery that they observed. Investigating groups larger than dyads, (Moussaid et al., 2010) reported a linear decrease of speed with growing group size.

**Figure 2** presents empirical measurements of group speed for groups including 1 (single individuals) to 4 members, collated from diverse published works. (Of course, we do not make any claim of exhaustiveness in the content of the figure). The first observation that one can make is the systematic reduction in speed as the number of members grows from 1 to 3, and the less clear change between 3 and 4-member groups, possibly partly owing to statistical uncertainty.

Some convincing explanations have been put forward to rationalise the observed reduction in speed. They involve the '*need to maintain spatial cohesion to communicate*' (Gorrini et al., 2016), the *'cognitive load'* associated with the intra-group social interactions (Zanlungo et al., 2014), the deceleration to adjust one's head orientation to facilitate eye contact and communication, and possibly the 'non-aerodynamic' profile of some group shapes at high densities (Moussaid et al., 2010) (note that some of the foregoing explanations bear similarities).

While the trend of reduced speeds for larger groups systematically appears in **Fig. 2**, the variability between the different studies of the speed measured from one group size is conspicuous. Speed values for individuals range from 1.0 to about 1.4 m/s and the difference is even larger for triads, with values ranging from about 0.7 to 1.2 m/s. Similarly to the question of group configuration, part of these variations can be accounted for by a more careful empirical inspection of the groups from the standpoint of gender composition, age, height, and purpose. Indeed, male dyads and triads walk faster than their female counterparts (Zanlungo et al., 2017; Zanlungo et al., 2019). Groups with tall people walk faster, while those with elderly or children move at reduced speed, which is in line with (Gorrini et al., 2016)'s observation that ederly pedestrians walked 40% slower than other adult groups (Gorrini et al., 2016). Finally, groups of work colleagues move faster than leisure-oriented groups (Zanlungo et al., 2017).

**Figure 2**: Dependence of the mean speed of the group on its size, according to the publications listed in the legend. The error bars represent standard errors. For (Schultz et al., 2014), the central values are medians and the standard deviations σ have been inferred from the interquartile range (IQR) using $\sigma \simeq \frac{IQR}{1.35}$ under the assumption of normally distributed data.

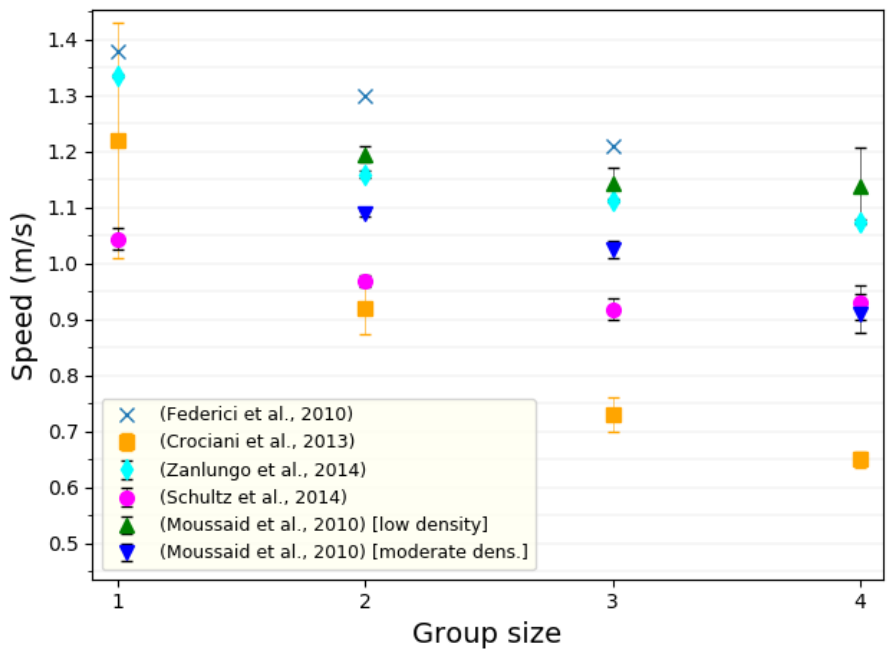

### *D. Automated detection and identification*

To conclude this first Chapter, we note that scientific endeavours to automate the detection of pedestrian groups are thriving, which is only partly caused by the Covid-19 related efforts to distinguish related people from non-related ones on videos, in order to verify compliance with the physical distancing guidelines.

Already in 2012, and following (Zou et al., 2011)'s focus on the statistics of private space in pedestrian groups, (Ge et al., 2012) proposed an automatic detection of groups, based on the criteria of (McPhail & Wohlstein, 1982). They found that pedestrians who are walking side-by-side are more likely to form a group compared to pedestrians who are walking at similar distance but in front-to-back configuration. (Khan et al., 2015) resorted to computer vision for the same task, using only the start and end points of trajectories, instead of the full ones, to classify agents into groups. More recently, (Pouw et al., 2020) proposed an algorithm to cluster the network of interactions, weighted by mutual distances and contact times, so as to identify family or group clusters.

# II. Impact of social groups on the collective dynamics

After the identification of social groups and the exploration of their 'microscopic' properties conducted above, this Chapter is aimed at gathering empirical data about the impact of groups on crowd dynamics.[3] Some (limited) numerical results will also be presented, but their relevance will be put to the probe by comparison with experimental data as much as possible.

### *A. Unidirectional flows*

The fundamental diagram associated with unidirectional flow is naturally a feature of paramount importance for pedestrian studies (as it is for vehicular traffic): It is central for the dimensioning analysis of facilities and, in particular, corridor widths so that they meet desired standards either in operational or in emergency modes. Therefore, it features among the criteria propounded by (Köster et al., 2014) for the validation of crowd models with social groups.

Several relations between the specific flow $J$ and the density $\rho$ have been put forward, notably that of Weidmann (1993), on the basis of empirical studies or controlled experiments. However, in the past decade researchers have started to wonder about the influence of social groups on these results. Somewhat surprisingly, while the reduced speed of social groups in free flow is well established (see **Fig. 2**), small groups may have only little impact on the fundamental diagram, at least at medium to high densities (higher than 1 ped/m²). Indeed, the latest experiments by (Hu et al., 2021) found that, at densities between 1.25 and 4 ped/m², crowds made of individuals ('atomic') did not flow considerably faster than crowds made of dyads, with the difference between the two situations hardly visible in the midst of the experimental variability. In fact, atomic crowds even seemed to display a slightly more pronounced slowdown as the density increased, possibly owing to the higher capacity of groups to adapt their configuration. That being said, note that this result is at odds with the output of the discrete model of (Crociani et al., 2017b), predicting a lower flow in the presence of dyads, and the experiments with counterflow of (Crociani et al., 2017a) [see Fig. 6 of that reference].

---
[3] Collision avoidance between groups will be addressed in **Chapter III**.

## B. Multidirectional flows

At low density, one expects to recover the situation of free flow, where groups are slower. This expectation is confirmed by passive observations of popular Fairs in Germany (Oberhagemann, 2012), in which one should however underline that the flow was multidirectional, and not unidirectional. These empirical results are particularly interesting because the dependence of speed on *both* the group size and the local density is reported. In **Fig. 3** we plot the data tabulated in (Oberhagemann, 2012) to show these dependencies. It is remarkable that the discrepancy of speed depending on size at low densities tends to shrink at higher density.

These findings are not corroborated by the controlled experiments of bidirectional flow performed by (Crociani et al., 2017a), in which similar speeds are reported for sparse crowds, whether atomic or made of groups; as the density grows the latter exhibit a more pronounced reduction[4] in speed, suggesting that the peak flow may be lower too. However, in these experiments, only groups of size 2 (i.e., dyads), consisting of participants asked to *try to keep close to one another*, were considered and these dyads were mixed in the crowd with individuals, which may explain part of the difference. Indeed, the difference in speed between individuals and dyads is less than **0.2m/s in Fig. 3**, so that the flow of a crowd mixing these two constituents is not expected to differ much from that of an atomic crowd.

---

[4] A numerical model by (Zanlungo et al., 2020) also finds a stronger speed reduction in crowds involving groups, except at very high densities.

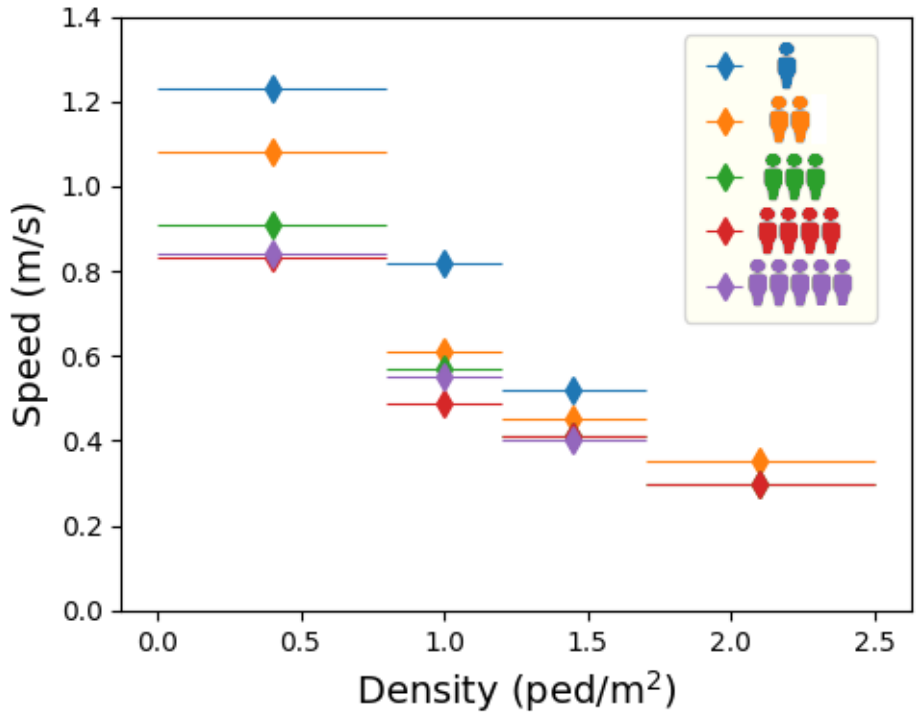

**Figure 3**: Group speed as a function of the group size and the crowd's density, plotted using the data of (Oberhagemann, 2012), Tab. 1, p. 20.

## *C. Bottleneck flows and emergency evacuations*

Evacuation is obviously a central concern for safety science and for the design of public facilities, which must ensure the possibility of a safe and efficient evacuation in the event of an emergency. Casualties are still grieved every year because of too lengthy or hampered evacuations when a safety hazard took place.

Despite this importance and the numerous studies dedicated to the topic, the influence of social groups on evacuations remains largely uncertain and controversial, with some apparent contradictions in the literature between reports of positive effects (Von Krüchten et al., 2017) and indications of detrimental ones (Bode et al., 2015). At least two factors can explain the current state of things.

*(1) Complexity of the evacuation process*

The first one is that an evacuation is in fact a complex process, consisting of different stages (Nicolas et al., 2016; Bernardini et al., 2019): (i) a pre-movement stage, when the decision is taken by people to start evacuating, (ii) a movement stage, where an escape route is selected (Hassan, 2013), either individually or collectively, and followed, (iii) possible delays at bottlenecks, whose limited capacity limits the outflow of the crowd. Regarding the pre-movement stage (i), controlled experiments suggest that social groups may take longer to respond to an alert (Bode et al., 2015). Nevertheless, it is not clear to what extent these results obtained in a specific setting with a limited number of participants can be generalised to real emergency scenarios.

The choice of an escape route [point (ii)] is influenced by different factors (Hassan, 2013), including the visibility conditions, knowledge of the premises and the locations of the exits, psychological state of the crowd, and the presence of leaders and social groups (Li et al., 2020). Often, there is an interplay between these factors, so that their individual influence is hard to disentangle. For instance, controlled experiments with participants wearing partly opaque eyepatches suggest that maintaining social groups may delay the evacuation under normal visibility conditions and that suboptimal structures (such as large groups walking side by side) are then seen. To the opposite, under reduced visibility groups had a positive effect, with an enhanced tendency to follow one another towards the exit, often by means of physical contact (Xie et al., 2020). Note in passing that the ability to infer somebody else's movement direction improves substantially with experience (Mojtahedi et al., 2017), so the inference might be more successful for socially related people. The network of pre-existing social relations helps understand or predict the formation of groups or isolated evacuations in the event of an evacuation (Li et al., 2020); pedestrians will tend to follow people whom they know, or who have great leadership or are the most talkative. Guidance and exchange of information between acquaintances towards a less congested exit path (e.g., staircase) have also been witnessed during evacuations (Han & Liu, 2017; Bernardini et al., 2019). In a similar context of study, simulations suggest that leaders can efficiently guide the crowd in poor visibility conditions (only a small fraction of such leaders is needed for the evacuation of large crowds), but their impact becomes negative when visibility is satisfactory (Ma et al., 2016).

Finally, the delays possibly incurred at bottlenecks (doorways, narrowings, etc.) have been the focus of considerable research, but the effects of social groups in this regard remain debated. One should recall that, for the most competitive egresses, arches of pedestrians will form at the exits and may produce transient clogs, which can delay the egress (Pastor et al., 2015) but also lead to the build-up of high pressures. The formation of these arches varies with the body size of pedestrians, their behaviour, and the dimensions of the exit (Yu et al., 2005). Body size and exit width are to be taken into account relative to each other: Arches form when the exit cannot accommodate many pedestrians passing simultaneously. Behavioural effects are also intuitive: As selfish actions and vying behaviour are witnessed, pedestrians become aggressive and try to squeeze past those in front. When an individual collides with others, they choose the temporarily accessible route nearest to them, and squeeze past. However, they cannot squeeze through the gaps of preceding pedestrians if these are too narrow. These physical interactions among people cause arching and clogging, and jams build up around the exit.

Running counter to this effect, it has been suggested, on the basis of controlled experiments with crowds of 32 to 46 pupils, that large social groups may reduce the evacuation time thanks to the enhanced capacity to self-order near the exit; the formation of queues does indeed mitigate the foregoing issues (jams) (Von Krüchten et al., 2017). The effect of dyads, however, is seen as detrimental in (Crociani et al., 2017b), owing to the resulting slower walking speed and lower pedestrian flow. In the light of experiments involving 12 pedestrians in a square room (10x10m) with six exits, (Bode et al., 2015) also report a longer time to move close to the exits for members of groups, compared to individuals.

*(2) Complexity of human behaviour*
The second factor that may explain the unclear state of things lies with the complexity of human behaviour. A dedicated review (Zhu et al., 2020) highlights the critical importance of human behaviour on the outcome of building emergencies, both in relation with the choice of an escape route and in the interactions with other evacuees. Diverse manifestations of social attachment that have been observed to influence emergency response are reported in (Bernardini et al., 2019). Besides, human behavior is always correlated with the state of their social environments, building environments, and emergency situations (Thalmann et al., 2000; Zhao et al., 2020; Zhu et al., 2020).
In this respect, too, social groups may have opposite effects (Ma et al., 2017). Indeed, the cooperative behaviour of people within small groups can favourably affect the evacuation time (Ma et al., 2017). On the contrary, the lower speed of groups, especially those formed in the presence of intimate social relations, may slow down the overall evacuation and disturb the crowd flow, and the internal cohesion of groups may come at the expense of less attention to other people, resulting in more vying behaviours between groups (Von Krüchten et al., 2017).

## *D. Miscellaneous*

Beyond evacuation issues, the detection and the analysis of pedestrian groups have found previously unexpected applications in the context of the Covid-19 pandemic. Indeed, social distancing guidelines have been enforced by and large in many countries in order to limit the risks of transmission of the SARS-CoV-2 virus. Commercial software solutions (notably based on machine-learning) have been proposed to automatically control the abidance by these prescriptions in large facilities. However, this requires identifying social groups, because social distances need not be respected within tight groups such as a parent and their children or a couple. Accordingly, in their recordings on a platform of the Eindhoven train station, (Pouw et al., 2020) first identified social groups and then measured separation distances which were found to be larger than before the enforcement of the rules.

Beyond these observations, groups also impact the model-based assessment of the actual risks of viral transmission in pedestrian crowds. Indeed, if one pedestrian on a street or in a public facility happens to be infected and contagious, the risks of transmission to other pedestrians should be modulated according to their relation with the index patient: If they are members of his or her group, it is very likely that they have already interacted with the index patient in situations that raise higher risks (e.g., at home), whereas unrelated people are only exposed in the observed street or pedestrian facility. This distinction led (Garcia et al., 2020) to exclude fellow group members in their assessment of the risks of new infections amid outdoor crowds, which were evaluated using spatially resolved transmission models. It seems to us that this 'group factor' deserves more attention in Covid-related studies of pedestrian dynamics, where it has often been overlooked so far.

## III. Methods to model groups in numerical models

The previous Chapters have shed light on some prominent characteristics of social groups in crowds and the (established or still hypothesised) impact of these group structures on the collective dynamics. How, and to what extent, are these features captured in numerical models for pedestrian dynamics? By addressing these questions, this section aims to light the path for prospective modellers in a dither. Therefore, we shall keep clear from the desire to exhaustively enumerate all previous works or detail their implementation in depth (which can be found in the references). Instead, we shall endeavour to delineate the main modelling approaches, their mutual connections, as well as their successes and limitations.

We have identified three key questions regarding the modelling of groups,
1) How is the cohesion, or compactness, of the group maintained in the simulations?
2) How is the configuration of the group, notably in terms of relative positions, replicated?
3) At a more strategic level, how are specific behaviours (expected from group members in some situations) implemented?
As will be shown, the answers largely depend on whether the simulations consider a leader guiding each group or not. In the following, we address these questions first in the case of lattice-based models, widely known as cellular automata (CA). Secondly, turning to continuous models, we will see that the strategies often mirror those introduced for CA, but raise particular challenges.

## A. Lattice-based models (cellular automata)

CA are widely used in the field of pedestrian dynamics (Burstedde et al., 2001; Maury and Faure, 2018), perhaps in large part because of their easy implementation and the fewer technical issues and unexpected behaviours that they induce, compared to their continuous counterparts, at the expense of less versatility. In these lattice-based models, the geometry of space is discretised, so that agents evolve on a checkerboard, hopping from cell to cell (or from a set of cells to another one, in the more recent versions featuring a *finer meshgrid*). These discontinuous hops in space entail a discretisation of time as well. Movements are thus jerky, which opens different possibilities for the update sequence: Do all agents try to move at the same time (*parallel update*) or do they move one by one (*sequential update*), and in the latter case always in the same order (*frozen*) or not?

When her turn arrives, an agent may try to hop to a neighbouring cell that is deemed attractive, from the vantage point of a generalised cost, or 'floor field', $V_i = V(x_i, y_i, O_i, ...)$ which will notably depend on the cell coordinates[5] and current occupancy $O_i$[6]. With these notations, a deterministic evolution leads to the selection of the neighbouring site $j$ with lowest $V_j$; nonetheless, stochastic rules are often preferred, where site $i$ is selected e.g. with a probability

$$p_i = \frac{e^{-V_i}}{\sum_j e^{-V_j}}$$

This discrete-choice model has the advantage of being familiar both to physicists, who interpret it as a Boltzmann distribution with $k_B T = 1$, and to economists, who regard it as a multinomial logit model with utilities $U_i = -V_i$. Note that the parallel with discrete-choice models has been pushed further by developing step-based models with utility functions that are actually specified and calibrated to minimise the difference with empirical data (Robin et al., 2009), as is commonly done in transportation theory.

### 1. Group cohesion

With this succinct presentation in mind, there is a straightforward pathway to modelling the cohesion of groups (*question 1 above*), especially if there is a group leader. It simply consists in lowering the floor field $V_j$ of cells in the vicinity of the group leader to promote the aggregation of other members, much in the same way as does an exit. However, unlike the exit, the leader and the group members move during the simulation, so the induced floor field component is called *dynamic* (in contrast to the *static* floor field generated by the geometry).

---

[5] In the case of an egress, agents are attracted to the exit, as though the floor were tilted downwards in that direction, with lower $V$ values.
[6] In general, only one agent is allowed per cell, in which case the cost $V$ tends to infinity if the cell is occupied.

This strategy is best illustrated by mentioning a few examples. (Lu et al., 2016) consider a dynamic part of the floor field $V_i = k_d \cdot d_i$, where $k_d$ is a positive constant and $d_i$ is the distance between the agent under study and her group leader. Alternatively, instead of penalising the distance $d_i$ to the leader, the utility function may penalise the area $A_i$ of the convex polygon formed by the group members (Crociani et al., 2018), which is better suited to groups with no hierarchy. While a more complex utility function is used in (You et al., 2016), the preservation of the group cohesion is more trivial, in that each group is rigid: it moves as a rigid body, whenever this move is possible for all members.

In any case, the movement of the group (i.e., of its centre of mass) in the infrastructure then largely relies on the group leader and the rules governing its evolution; the many-body group problem thus boils down to a one-body problem.

*2. Group shape*

If there is no hierarchy in the group, members may be attracted to the centre of mass of the group, or uniform motion may be imposed on them under the control of a particular group member. For instance, (You et al., 2016) impose that the member who gains most thanks to the move (i.e., the member who has the option of motion whose generalised cost is minimal) determines the step choice for *all* members. The other group members then follow suit with the same step, which immediately solves the question of how to maintain the group structure (*question 2 above*). Indeed, since all group members always make identical steps, the shape of the group naturally remains constant.

A less rigid strategy to promote coherent group motion is to increase the utility of step choices that mirror the leader's stepping direction, viz., $V_i \to V_i - k_i$, where $k_i > 0$, if the agent steps in the same direction as the group leader. This maintains the spatial structure of the group at low density, but not at higher density (Lu et al., 2016). This promotion of alignment is of course reminiscent of the flocking term used in continuous physical models (Vicsek et al., 1995), as we recall below. A related strategy was employed in the leader-follower model of (Robin et al., 2009), insofar as each follower selects a leader among the 'similar' neighbouring individuals and tries to match his velocity.

Still, the relaxation of the assumption of a rigid group shape gives rise to different issues, even when the groups only comprise two members. In particular, conflicts over cell occupation may artificially arise within a group, when the members aim for the same cell. (Crociani et al., 2018) remedied several of these issues, at the expense of a fairly complicated utility function, and calibrated and validated the resulting CA against controlled experiments, with a particular focus on dyads.

*3. Specific behaviour*

In some circumstances, being part of a social group may affect higher-level ('strategic') decisions, which will then differ from those of single walkers, whether it be by entailing differences in behaviours, in the priority of tasks to complete, or in the routes to follow. These differences should be suitably taken into account in models.

For instance, the evacuation of a pediatric hospital involves groups of parents and children and the former are expected to take the latter in their arms to speed up the evacuation. Therefore, to describe this situation, (Li et al., 2021) allowed a child to share the same cell as one of his or her parents in their CA; group members are thus allowed to share a cell.

**Modelling of evacuations**

The special case of evacuation models deserves a whole paragraph of its own, because it has attracted much attention. As explained in the previous Chapter, an evacuation is a multi-stage process, with different levels of decision-making (Murakami et al., 2019), from the decision to evacuate to the choice of an exit, and bottlenecks possibly present a risk in emergency conditions. To handle the latter, the modeller needs to (i) describe the geometry of the bottleneck and clarify the emergency situation, (ii) insert the right number of people in the room and consider its degree of congestion, (iii) model the behaviour of people ahead of, and at, the bottleneck. Social groups are then inserted into the ensuing framework. As an illustration, in (You et al., 2016), the will to egress is described by the floor field

$$V = \alpha\, Distance + \beta\, Density + \frac{\delta}{\overline{v}_{visible}},$$

where *Distance* refers to the distance to the exit, while *Density* and $\overline{v}_{visible}$ are the density and speed of pedestrians in the cone of vision, and α,β, and δ are positive parameters. Groups are then described as rigid structures moving together in the direction of the member who gains the most (lowest *V*) through the move. **Table 2** summarises the characteristics of some of the models used to study emergency evacuations numerically. In these simulations, the presence of social groups is generally found to have an effect on the evacuation time (Köster et al., 2015; You et al., 2016; Crociani et al., 2019); the effect is often adverse due to the structural rigidity of groups.

However, in addition to these effects at the bottleneck, the presence of social groups may alter the evacuation behaviours. In particular, in emergency evacuations, we expect people with strong social ties to be inclined to wait for one another and possibly backtrack to search missing members (consider the case of parents moving away from the exit to look for their missing children). These specific behaviours are usually implemented via sets of *ad hoc* algorithmic rules applying to group members; for instance, the 'wait + backtrack if a group member is missing' was implemented in the cellular automaton of (Yang et al., 2005). For complex sets of rules, it may be suggested to build action decision-trees, specifying different actions depending on a series of yes/no conditions, as was done by (Ta et al., 2017) in an evacuation context.

**Table 2 - Characteristics of social group behaviour during emergency evacuation found in a selection of simulations**

| Reference | Experimental setting | | | | Social group behaviour |
|---|---|---|---|---|---|
| | **Model** | **Crowd size** | **Density** | **Flow[7]** | |
| Xie, R. et al, 2016 | Dynamic force model | NA | Low and moderate | M | Group aggregation phenomenon, especially the tendency for pedestrians to stay closer to socially intimate group members instead of simply moving to the 'group center' |
| Zhang, H. et al., 2018 | SFM[8] | 100 - 300 | Low to high | U | Modified SFM in which leaders are selected at central positions in the initial stage and then head for the exit; group members have a desired velocity directed towards their leader; isolated people may join groups. |
| Şahin, C. et al., 2019 | Multi-agent model | 60 - 160 | High | U | Multi-agent model with mobility, health, age, stress and panic attributes. Simulated on Unity platform, with an undescribed dynamical model. |
| Chu, M.L. and Law, K.H., 2019 | Hybrid | 13,200 | Moderate to high | M | No actual social group affiliation, but simulated stewards may manage crowd evacuation, improving the evacuation efficiency. |
| Han, Y. and Liu, H., 2017 | SFM | 50 - 250 | High | M | Modified SFM to include (i) an information transmission mechanism (notably within groups) that allows agents to choose an exit based on neighbours' information, instead of the nearest one; (ii) leaders or guides; (iii) improved collision avoidance. |
| Liu, B. et al., 2018. | SFM | 100 - 300 | High | M | The SFM is supplemented with an attractive force between agents, whose strength depends on the mutual relationship. An exit is chosen by the group leader. Both initial position and target goals are defined based on empirical video observations |

---

[7] **U** stands for Unidirectional / Unique exit and **M** stands for Multidirectional / Multiple exits.
[8] SFM: Social Force Model

| Zhao, X. et al., 2020. | Random forest | 569 | NA | M | Focus on the pre-movement phase. Machine-learning of the decision-making time based on data from drills; the influence of social factors is highlighted. |

## B. Continuous models

*1. Group cohesion*

We now turn to continuous agent-based models. To clarify their relation with CA, one should first realise that physical equations in continuous space are often amenable to a numerical resolution on a discrete lattice. A celebrated example is the recourse to a lattice-Boltzmann method to solve the Navier-Stokes equation, the main equation of fluid dynamics: Instead of solving the partial differential equations of the Navier-Stokes equation, one may rather consider the exchange of fictitious 'fluid particles' on a lattice, which gives the same solution in the hydrodynamic limit (at least in some regimes), as can be shown theoretically by the so-called Chapman-Enskog expansion. In a similar vein, although admittedly without a rigorous demonstration of equivalence, CA implementations of the originally continuous Social Force Model of (Helbing and Molnar, 1995) have been proposed; see e.g. (Li and Liu, 2017). This very popular model is premised on Newton's equation of motion for the agents, viz.,

$$m\ddot{r}_i = m\frac{v_i - \dot{r}_i}{\tau} + \sum_{j \neq i} f_{j \to i} + f_{wall \to i} + \xi_i,$$

where $m$ denotes a mass, $\tau$ is a characteristic reaction time, $v_i$ is the desired velocity of the agent, $f_{j \to i}$ and $f_{wall \to i}$ are binary forces exerted on $i$ by other agents $j$ or obstacles, and $\xi_i$ is a stochastic term. Note that, here, $f_{j \to i}$ includes both medium-ranged social interactions and short-range contact interactions. To further the analogy with physical systems, the binary forces may be assumed to derive from a pair potential $V_{ij}$, viz., $f_{j \to i} = -\frac{\partial V_{ij}}{\partial r_i}$, the sum of these forces thus giving a contribution $\sum_{j \neq i} V_{ij}$ to the dynamic floor field, while the stochastic term $\xi_i$ accounts for the stochastic nature of the CA rules.

If one sticks to this analogy, the attractive interaction term used in CA to enforce the cohesion of groups is equivalent to *radial* forces $g_{ij}(r_{ij})n_{ij}$ directed towards $j$ in the continuous equation of motion for agent $i$, where $g_{ij}$ is a function that depends on the distance $r_{ij}$ between $i$ and $j$ and $n_{ij}$ is the unit vector binding these two agents. This approach relying on radial cohesive forces is indeed adopted in (Li and Liu, 2017; Liu and Li, 2018), where $g_{ij}(r) = w_{ij} \exp(\frac{r-d_{ij}}{B_i})$ if the spacing between group members $i$ and $j$ is larger than the characteristic distance $d_{ij}$ and zero otherwise (there is no pull if the agents are already close to one another). Here, $B_i > 0$ and the $w_{ij}$ weights measure the affinity between $i$ and $j$.

In the presence of a group leader, one may settle with a unique radial force per member, which pulls her towards the leader, instead of having binary interactions between all group members. Alternatively, this force can also be directed towards the centroid of the group (Singh and Drury, 2009).

As in CA, this attractive term can be either substituted by, or complemented with, a term promoting alignment of the individual velocities of group members (Qiu and Hu, 2010; Chen and Wang, 2020). This term is often expressed as the average of the velocities of the $N$ group members, $\frac{1}{N}\sum v_j$, up to a multiplicative coefficient, with a direct inspiration from the flocking term in Vicsek's model for bird flocks (Vicsek et al., 1995; also see Reynolds, 1987).

While these simple approaches promote the clustering of related pedestrians and their walking together, they may fall short of capturing some notable features of pedestrian groups.

In this regard, the first issue is actually not specific to pedestrian crowds, but also applies to a broader range of active systems. Indeed, physicists have come to realise the central role of perception to maintain the cohesion of groups of self-propelled particles (Lavergne et al., 2019). Importantly, perception differs from a physical interaction in that it may be, and often is, asymmetric, that is to say, $f_{i \to j} \neq - f_{j \to i}$. This asymmetry notably undermines the aforementioned assumption of forces deriving from a global energy potential, in the equations of motion, for a broader repertoire of interactions are then allowed.

This first crack in the analogy between the equations of motion of physical systems and those aimed at describing the dynamics of pedestrians actually points to larger rifts, which were less apparent in rule-based CA. These rifts only become deeper due to the fact that the motion of decision-making entities does not result only from a physical layer (Nicolas, 2020). In particular, their behaviour (understood as their response to their particular environment) may be governed by perception and emotion, as well as their memory. The schematic model of (Thalmann et al., 2000), represented in **Figure 4**, shows the hierarchical level that will be performed in sequence sometimes or concurrently to determine actions of various complexities. The specification of a model will require describing the interactions between people and groups of agents, but also those with static objects in the spatial layout, thus calling for the implementation of mechanisms of locomotion, collision avoidance, and navigation in an iterative behavioural loop. As stated by (Kleinmeier et al., 2020), '*humans use perception, cognition and a repertoire of different behaviours. They also interact*'.

In the context of walking in an outdoor setting, the influence of subjective perception on the choice of a trajectory (for instance, avoiding regions with threatening people) can be modelled with a perception-dependent repulsive force in the equations of motion (Colombi and Scianna, 2017). In the context of an evacuation, it can prompt a variety of possible actions summarised in a decision-tree (Ta et al. 2017). Such extensions of physical models may give access to a more realistic description. On the flip side, they extend the range of possibilities and free parameters and may thus obfuscate the global picture or lack generality.

**Figure 4:** Behavioral model (Thalmann et al., 2000): **a)** Sketch of the principle, **b)** Behavioural loop iteration in modelling the dynamic and static objects in a spatial layout

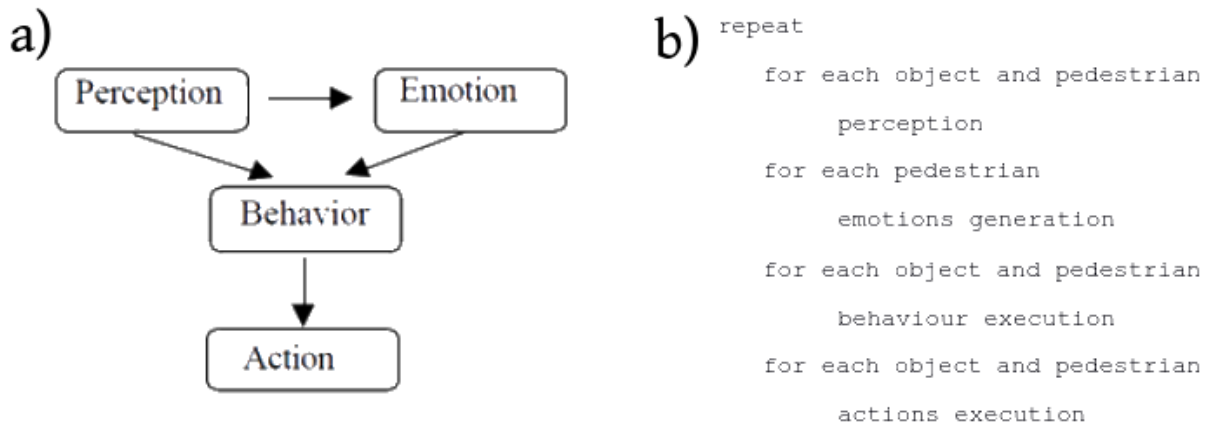

*2. Group shape*

The second crack that opens up if one settles with simple attractive and flocking forces lies with the faulty group shapes that are then obtained. Indeed, these forces do not give rise to the stable structures adopted by small groups in reality and exposed in **Section IB**. This deficiency, in turn, is likely to affect the interactions between the group and the rest of the crowd and thus possibly alter the global flow properties.

Several strategies have been proposed to overcome it.
(Qiu and Hu, 2010) opted for a refined description of the intra-group relationships (in addition to also including interactions between groups, mediated by their leaders). More precisely, they introduced a matrix of intra-group weights, that is here denoted by $w_{ij}$ and that measures the relationship between group members *i* and *j*. The aforementioned attractive ('aggregation') and flocking ('following') forces are then specific to each member *i* in that their components for the interaction between *i* and *j* are weighted by $w_{ij}$. Different choices for the $w_{ij}$ lead to either linear groups (abreast walking), or compact groups, or follow-the-leader patterns. Earlier research (Singh & Drury, 2009) had considered an exclusive interaction with one neighbour (for followers), which comes down to setting most intra-group weights $w_{ij}$ to zero, and no flocking term; but this enhanced simplicity had come at the cost of being able to reproduce fewer possible configurations.

A distinct strategy, applicable to small groups, does not require the definition of an intra-group matrix, but rather consists in adding new interaction terms within the group. In particular, in their extension of the Social Force Model, (Moussaid et al., 2010) supplemented the aforementioned attractive force (and a repulsive one to prevent overlaps) with a 'visual' force $f_i \propto - \alpha_i v_i$ that strives to minimise the angular difference $\alpha_i$ between the walking direction (i.e., the velocity $v_i$) and the gazing direction, which is directed towards the centre of mass of all *other* group members. The underlying idea is that pedestrians are eager to decelerate to be able to see the rest of the group without the discomfort of excessively turning their heads. This addition successfully captured not only the reduction of speed as the small group gets larger, but also the experimentally observed deviations from abreast (side-by-side) walking for small groups, i.e., the **V**-shape of triads and the **U**-shape of 4-member groups.

A related idea was propounded by (Zanlungo et al., 2014), who insisted on the fact that the force term at the origin of the group structure need not be symmetric (or `reciprocal') between agents *i* and *j*, unlike Newtonian forces. The authors chose to express this structuring term as the derivative of a discomfort potential $D(r, \theta)$ that depends on the distance *r* between the neighbours and on the discomfort angle $\theta$ between the direction of gaze (assuming they look at one another) and the walking direction. The different weights bestowed on one's own discomfort angle and that of one's neighbour break the reciprocity of the forces and, incidentally, also account for the reduction of speed as the group size increases. If the discomfort potential $D(r, \theta)$ only acts on first neighbours (and not on all other group members, as was done in the previous works), this method is able to quantitatively capture the average configurations observed for dyads and triads.

*3. Collision avoidance and virtual crowds*

Of course, the foregoing force terms designed to structure the shape of the group can be supplemented with other terms describing group behaviours, such as a seeking force which drives members towards forlorn companions. Quite generally, beyond the interactions within each group, the interactions between distinct groups need to be addressed: How do social groups avoid colliding with each other, while maintaining or adapting their structure?

The authors of (Huang et al., 2018) suggest that collision avoidance between groups consists of catch-up avoidance and face-to-face avoidance. The catch-up avoidance is the movement pattern whereby the rear groups will try to catch up with the front group as they move in the same direction. On the other hand, face-to-face avoidance is the movement pattern that occurs when groups moving in opposite directions come into one another and may block one another; this will cause holistic ('go-around') avoidance or interspersed ('go-through') avoidance. In the former, the group will maintain its compactness and avoid the other group as a whole, whereas interspersed avoidance allows outsiders to cross the group. Collision may also occur when members of a group are moving together towards the same point, guided by distinct leaders; groups larger than 3 will then often split into subgroups with 3 or 2 people as the front subgroup and the rear subgroup (Xi et al., 2014). During the movement process, there will be no direct interaction between the front and rear subgroups, except when their spacing exceeds a typical distance around 2 meters (Huang et al., 2018), at which point the rear subgroup will try to catch up with the front one.

In the case where collision avoidance is performed by a single individual facing a social group, the question had been addressed previously by (Bruneau et al., 2015). Using virtual reality experiments, they showed that participants tend to 'go through' large sparse groups, whereas they will 'walk around' small dense groups. For a circular group of radius 3 meters, the proportion of 'go-through' moves surges from around 20% to 100% when the interpersonal distance within the group increases from 1.1m to 2.3m. The extreme cases are well rationalised by a principle of minimum energy which compares the cost of a test *go-through* trajectory with that of a *walk-around* path. Unlike most of the aforementioned works, here, the trajectories were not computed with a force-based model, but with a velocity-obstacle model. In this branch of continuous models, for each agent, the space of possible velocities is explored in search of the best one in terms of an objective function, after excluding all velocities that are expected to lead to a collision (over a finite time horizon); the latter are called velocity-obstacles. (Ren et al., 2017) put forward a general-purpose method to introduce various types of groups within these models. It simply consists in further constraining the explored velocity space by barring velocities that drive an agent too far from her group neighbours at the next time step; a subtlety is that not all group members are regarded as neighbours, but only a subset of them defined by physical proximity and social relations. Finally, the explored velocity space may be overlaid with additional costs to favour specific group formations.

Quite interestingly, the foregoing endeavours were not undertaken primarily to understand the actual dynamics of crowds, but to animate virtual crowds e.g. in video games. In this context, the motion of groups, including their shape, was still deemed visually unrealistic by (Rojas & Yang, 2013). These researchers wanted to remedy this issue and also to allow the group structure to evolve smoothly over time. Two solutions were identified for this purpose, either by considering that group members form an articulated chain whose bonds can be adjusted, or by interpolating between predefined structure templates. On a more historical note, many of the mechanisms discussed above to maintain the cohesion of groups can in fact be traced back to a seminal work by C.W. Reynolds (Reynolds, 1987) dealing with the animation of bird flocks; the virtual counterparts of birds were 'boids', artificial creatures obeying rules that depend on their local environment. The three major rules governing their behaviour were indeed to (i) steer to avoid collisions, (ii) match the velocity of the neighbours, (iii) steer towards the centroid of nearby boids ('flock centering').

We would like to conclude by mentioning another line of modelling which was propounded by the same author but which has not been taken up in recent research, as far as we know. It is premised on the idea that, rather than being written by hand e.g. as an interaction term, the formation and coordination of groups could emerge algorithmically as the fruit of an evolutionary process aimed at fulfilling a given function. Thus, (Reynolds, 1993) simulated artificial agents (then called `critters') whose goal was to avoid collisions between themselves and stay clear of a predator keen on isolated stragglers. The programme encoding the behaviour of the critters was allowed to evolve with a genetic algorithm and it ultimately led to some degree of group motion. We are not aware of any similar endeavours in relation with pedestrian groups.

**Table 3** Summary of some of the reviewed pedestrian simulation models with social behaviour parameters.

| Interaction Behaviour Parameters | Setting in Simulation | Reference | Type of Simulation (Discrete or Continuous) |
|---|---|---|---|
| - Avoiding behaviour among different social groups.<br><br>- Coordinate behaviours among | - The group members will stick together and implement holistic avoidance. | (Huang et al., 2018) | Continuous |

| subgroups that belong to one social group. | - The rear subgroup will always try to catch up with the front subgroup. | | |
|---|---|---|---|
| - Avoiding behaviour among different social groups.<br><br>- The distance of the intra group pedestrians. | - Pedestrian crowd crossing a square place from different directions.<br><br>- The distances and angles between the intra group members are based on the measurement method by (Moussaïd et al., 2010) | (Prédhumeau et al., 2020) | Continuous |
| - Intra group structure.<br><br>- Inter group relationship. | - Agent based simulation with random movement, obstacle avoidance, and maintaining group ability. | (Qiu & Hu, 2010) | Discrete |
| - Collision avoidance between groups.<br><br>- Collision avoidance within a group. | - Inter-group interaction in a narrow corridor.<br><br>- Fast moving group encounters a slow moving group heading in a similar direction. | (Karamouzas & Overmars, 2012) | Discrete & Continuous |

**Conclusion**

The proportion of pedestrians walking in a social group varies widely with the context, from a small minority in some pedestrian facilities to a large majority at fairs and popular events. Remarkably, regardless of the context, small groups display some robust features, in particular a tendency to walk abreast for dyads, in a **V**-shape for triads, in a **U**-shape for groups of 4 and 5 members, as well as a reduced speed on average compared to individuals. These microscopic structural features are now fairly well established empirically, including their dependencies on purpose, gender, etc. (Zanlungo et al., 2017, 2019, 2020).

At a more global scale, the impact of these groups on the collective flow properties is more controversial and may be sensitive to the specific conditions of observation. We firmly believe that a better distinction of the different levels of decision and stages of motion could clarify the picture, especially in the case of evacuation scenarios: these scenarios lie at the intersection of diverse complex problems, from the pre-movement behaviour to route choice and to congestion at exits and narrowings. A more systematic delineation of the (overarching) strategic level, the (medium-scale) tactical level, and the (fine-scale) operational level (Hoogendoorn & Bovy, 2004) could be of some avail for that purpose. These remarks are also applicable to controlled experiments, where a narrower focus on specific aspects of the process under study may afford more solid conclusions and tell apart context-dependent effects and generic ones.

In particular, it would be of paramount importance to understand under what conditions the collective flow properties are affected by the presence of social groups at the operational level and, when this is the case, if these influences can be subsumed into aggregate variables (for instance, by adjusting the pedestrian density or the free walking speed depending on the crowd's composition) or if they hinge on a detailed description of groups. This question is naturally crucial for macroscopic models, which handle the crowd as a continuum and have remained largely impervious to the presence of social groups (hence their oversight in this review). But the question also impacts microscopic modelling, in that it determines how critical the modelling of these groups is for practical applications and how sophisticated it must be. Overall, at the strategic and tactical levels, in light of the complexity of human behaviour, further empirical validation is probably needed to ascertain the possible influences of the presence of social groups on the pedestrian dynamics, despite recent efforts to move in that direction (Bernardini et al., 2019).

At the operational level, 'microscopic' (agent-based) models could help rationalise the seemingly conflictual effects of social groups on the flow properties observed experimentally (Von Krüchten et al., 2017; Crociani et al., 2017b; Hu et al., 2021) by pinpointing the mechanisms at the origin of the observed trends. Moreover, they make it possible to explore scenarios beyond the ones that can (safely) be tested experimentally. Social groups have been integrated into these models by means of rules and strategies that differ considerably. These discrepancies are not a mere reflection of the gap between discrete models and continuous ones, which actually tends to shrink with the advent of multi-grid approaches in cellular automata and the development of continuous algorithms based on minimisations (Ren et al., 2017); they also exist within each category (ranging from the prescription of rigid group structures to groups that may split during encounters with other groups or obstacles). This diversity in modelling approaches may raise some questions about their realism. The development of validation procedures to prove the efficacy of the method to model groups (Köster et al., 2014; Crociani et al., 2018) should thus be encouraged. However, as for validation procedures of pedestrian dynamics software in general, these efforts are not devoid of difficulty: *general* validation methods should be strict enough (and probably stricter than they currently are) to avoid erratic behaviours in some circumstances, whereas a validation designed for a *specific* scenario does not warrant the applicability of the model outside the specific conditions of the observations that are at its basis.

Beyond the necessity to validate the *output* of the models, a more conceptual debate relates to the allowable *input* ingredients: Should the models rationalise the emergence of observed features associated with social groups starting from basic principles (Reynolds,

1993; Moussaid et al., 2010; Zanlungo et al., 2014; Bruneau et al., 2015) or is it acceptable to encode these features by hand? Naturally, the answer may strongly depend on the purpose of the model and might thus drive a wedge between distinct research communities. Au contraire, it seems to us that the topic of social groups is ideally suited to unite the video game community (and in particular the animators of virtual crowds) and the pedestrian dynamics community, much in the same way as pedestrian modellers and autonomous robot programmers joined forces to develop collision avoidance strategies. Along this vein, let us conclude this review by recalling that key ideas for the modelling of social groups can actually be traced back to efforts in the 1980s aimed at animating virtual flocks.

## Acknowledgements
We acknowledge funding from the French and Malaysian governments under the Hubert Curien Partnership France-Malaysia Hibiscus (PHC- Hibiscus) programme [203.PKOMP.6782005].

## Declaration of interest statement
The authors declare no conflict of interest.